\documentclass[times, 10pt,twocolumn]{article} 
\usepackage{latex8}
\usepackage{times}
\usepackage{epsfig}
\usepackage{graphicx}

\newcommand{\ie}{{\it i.e.,$\ $}}
\newcommand{\eg}{{\it e.g.,$\ $}}

\begin{document}

\thispagestyle{empty}

\setcounter{page}{1}
\twocolumn
\normalsize

\title{Byzantine Fault Tolerance for Nondeterministic Applications\thanks{
This work was supported in part by Department of Energy 
Contract DE-FC26-06NT42853, and by a Faculty Research Development award
from Cleveland State University.}}

\author{Wenbing Zhao\\
Department of Electrical and Computer Engineering\\ 
Cleveland State University, 2121 Euclid Ave, Cleveland, OH 44115\\
wenbing@ieee.org\\
}

\maketitle
\thispagestyle{empty}

\vspace{-0.1in}
\begin{abstract}
All practical applications contain some degree of nondeterminism.
When such applications are replicated to achieve Byzantine fault 
tolerance (BFT), their nondeterministic operations must be controlled 
to ensure replica consistency. To the best of our knowledge, only the 
most simplistic types of replica nondeterminism have been dealt with.
Furthermore, there lacks a systematic approach to handling common types of
nondeterminism. In this paper, we propose a classification of common
types of replica nondeterminism with respect to the requirement of
achieving Byzantine fault tolerance, and describe the design and
implementation of the core mechanisms necessary to handle such
nondeterminism within a Byzantine fault tolerance framework.
\end{abstract}

\vspace{-0.1in}
\noindent
{\bf Keywords}: Byzantine Fault Tolerance, Intrusion Tolerance,
Replica Nondeterminism, Security, Fault Tolerance Middleware

\vspace{-0.1in}
\Section{Introduction}
\vspace{-0.1in}
Today's society has increasing reliance on services provided over the
Internet. These services are expected to be highly dependable, which requires
the applications providing such services to be carefully designed and 
implemented, and rigorously tested. However, considering the intense 
pressure for short development cycles and the widespread use of 
commercial-off-the-shelf software components, it is not surprising 
that software systems are notoriously imperfect. 
The vulnerabilities due to insufficient design and poor implementation 
are often exploited by adversaries to cause a variety of damages, e.g., 
crashing of the applications, leaking of confidential information, 
modifying or deleting of critical data, or injecting of erroneous 
information into the application data. These malicious faults are often 
modeled as Byzantine faults.
One approach to tackle such threats is to replicate the server-side
applications and employ a Byzantine fault tolerance (BFT) algorithm as 
described in \cite{bft-osdi99, bft-osdi2000, bft-acm, base}.

Byzantine fault tolerance algorithms require the replicas to operate
deterministically, \ie given the same input under the same state, all
replicas produce the same output and transit to the same state.
However, all practical applications contain some degree of nondeterminism. 
When such applications are replicated to achieve fault and intrusion 
tolerance, their nondeterministic operations must be controlled to ensure 
replica consistency. 

To the best of our knowledge, only the most simplistic types of
replica nondeterminism have been dealt with under the Byzantine fault
model~\cite{bft-osdi99, bft-osdi2000, bft-acm, base}, which we term 
as wrappable nondeterminism and verifiable
pre-determinable nondeterminism. The former assumes that any
nondeterministic operations and their side effects can be 
mapped into some pre-specified abstract operations and state,
which are deterministic. The later assumes that any nondeterministic
values can be determined prior to the execution of a request, and
such values proposed by one replica can be verified by other replicas 
in a deterministic manner, and the values are accepted only if they are 
believed to be correct. 

The mechanisms designed to handle these types of nondeterminism 
either are not effective in guaranteeing replica consistency and/or are not 
effective in masking Byzantine faults, if the application to be replicated 
exhibits other types of nondeterministic behavior. For example,
many online gaming applications contain nondeterminism whose values 
(\eg random numbers that determine the state of the applications)
proposed by one replica cannot be verified by another
replica. It is dangerous to treat this type of nondeterminism
the same as the verifiable pre-determinable nondeterminism because a faulty
replica could use a predictable algorithm to update its internal state and 
collude with its clients, without being detected, which defeats the
purpose of Byzantine fault tolerance. As another example, multithreaded 
applications may exhibit nondeterminism 
whose values (\eg thread interleaving) cannot be determined prior to the 
execution of a request (without losing concurrency), which cannot be
handled by existing BFT mechanisms.

In this paper, we introduce a classification of common types of
replica nondeterminism present in many applications. We propose a set of 
mechanisms that can be used to control these types of nondeterministic 
operations. We also describe the implementation of the core mechanisms and 
their integration with a well-known BFT 
framework~\cite{bft-osdi99, bft-osdi2000, bft-acm, base}. Our performance 
evaluation of the integrated framework shows that our mechanisms only
introduce moderate runtime overhead.

\vspace{-0.1in}
\Section{Byzantine Fault Tolerance}
\vspace{-0.1in}
\label{backgroundsec}
This work is built on top of the BFT framework developed by Castro, 
Rodrigues, and Liskov \cite{bft-osdi99, bft-osdi2000, bft-acm, base}. 
We use the same assumptions and system models as those of the BFT framework. 
For completeness, we briefly summarize the BFT framework here.

The BFT framework supports client-server applications running in an 
asynchronous 
distributed environment with a Byzantine fault model, \ie faulty nodes
may exhibit arbitrary behaviors. It requires the use of 3f+1 replicas
to tolerate up to f faulty nodes. (In a recent publication~\cite{alvisi-bft},
Yin et al. proposed a method to reduce the number of replicas to
2f+1 by separating the executing and agreement nodes.)

\begin{figure}[t]
   \includegraphics[width=3.2in]{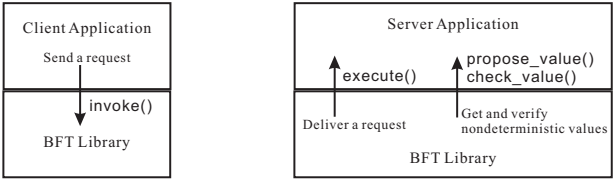}
   \caption{The positioning of the application and the BFT library, and the
core interfaces between the two components.} 
   \label{archfig}
\vspace{-0.25in}
\end{figure}

The BFT framework is implemented as a library to be linked to
the application code (both the server and the client sides), as shown
in Figure~\ref{archfig}.
In general, on the server side, we use the term {\em replica} to refer 
to the combined entity of the server application and the BFT library. 
On the client side, we use the term {\em client} to refer to the
combined entity of the client application and the client-side BFT library. 
Sometimes, however, it is necessary to distinguish the two parts explicitly. 
As shown in Figure~\ref{archfig}, the client-server application and the BFT 
mechanisms (residing in the BFT library) interact via a set of Application 
Programming Interfaces (APIs). The APIs contain a number of downcalls to be 
invoked by the application for a number of purposes, for example, to 
initialize the BFT library with appropriate parameters and callback functions, 
to send requests to the server replicas, and to start the event loop managed by
the BFT library. The APIs also contain a number of upcalls to be
implemented and supplied by the application, so that the BFT mechanisms
can deliver a request to the server application, retrieve
and verify nondeterministic values (if applicable), and retrieve and 
restore application state. Figure~\ref{archfig} includes a subset of
the APIs directly related to this work.

In the BFT framework, a replica is modeled as a state machine. The replica is
required to run (or rendered to run) deterministically.
The state change is triggered by remote invocations on the 
methods offered by the replica. In general, the client first sends 
its request to the primary replica. The primary replica then broadcasts the 
request message to the backup replicas and also determines the execution order 
of the message. All correct replicas must agree on the same set of request 
messages with the same execution order. In other words, the request messages 
must be delivered to the server application at all replicas 
reliably in the same total order. 

In the BFT framework, a very efficient Byzantine agreement algorithm,
often referred to as the BFT algorithm, is used to ensure the total ordering 
of the requests received from different clients. The normal operations
of the BFT algorithm involve three phases.
The first phase is called the pre-prepare phase, where the primary replica 
multicasts a \small\verb!PRE_PREPARE! \normalsize  message containing the
ordering information, the client's request, and the nondeterministic
values that can be determined prior to the execution of the request (if any) 
to all backup replicas. A backup replica then verifies 
the ordering information, the nondeterministic values, and
the validity of the request message. If the backup replica 
accepts the \small\verb!PRE_PREPARE! \normalsize message, 
it multicasts to all other replicas 
a \small\verb!PREPARE! \normalsize message containing the ordering 
information and the digest of the request message being ordered. This starts 
the second phase, \ie the prepare phase. When a replica has collected
2f valid \small\verb!PREPARE! \normalsize messages for the request from other 
replicas, it multicasts a \small\verb!COMMIT! \normalsize message. 
This is the start of the third phase. When a replica has received 
2f matching \small\verb!COMMIT! \normalsize messages from other replicas, 
the request message has been totally ordered and it is ready
to be delivered to the server application. This concludes the
third phase, \ie the commit phase, of the BFT algorithm.
In the BFT framework, all messages are protected by a digital signature, or
an authenticator~\cite{authenticator} to ensure their integrity.

\vspace{-0.1in}
\Section{Classification of Replica Nondeterminism}
\vspace{-0.1in}
\label{nondetsec}
We distinguish replica nondeterminism into the following three major
categories: 
\begin{itemize}
\item {\em Wrappable nondeterminism}. This type of replica nondeterminism
can be easily controlled by using an infrastructure-provided or 
application-provided wrapper function, {\em without explicit 
inter-replica coordination}. For example, information such 
as hostnames, process ids, file descriptors, etc. can be determined
group-wise. Another situation is when all replicas are implemented
according to the same abstract specification, in which case, 
a wrapper function can be used to translate between the local
state and the group-wise abstract state, as described in \cite{base}.

\item {\em Pre-determinable nondeterminism}. This is a type of replica 
nondeterminism whose values can be known {\em prior to} the execution of a 
request and it requires inter-replica coordination to ensure replica
consistency.

\item {\em Post-determinable nondeterminism}. This is a type of replica
nondeterminism whose values can only be recorded {\em after} 
the request is submitted for execution and the nondeterministic 
values won't be complete until the end of the execution. It also requires 
inter-replica coordination to ensure replica consistency.
\end{itemize}

In this paper, we will not have further discussion on the wrappable 
replica nondeterminism because it can be dealt with using a deterministic
wrapper function without inter-replica coordination, and also because
it has been thoroughly studied in \cite{base}. Instead, we will focus 
on the rest of two types of replica nondeterminism.

Based on if a replica can verify the nondeterministic values 
proposed (or recorded) by another replica, replica nondeterminism
can be further classified into the following types:
\begin{itemize}

\item {\em Verifiable nondeterminism}. The type of replica nondeterminism whose
values can be verified by other replicas.

\item {\em Non-verifiable nondeterminism}. The type of replica nondeterminism
whose values cannot be completely verified by other replicas. Note that
a replica might be able to partially verify some nondeterministic values
proposed by another replica. This would help reduce the impact of a faulty
replica.
\end{itemize}

Overall, our classification gives four types of replica nondeterminism of
our interest:
\begin{itemize}
\item {\em Verifiable pre-determinable nondeterminism (VPRE).} 
In the past, clock-related operations have been treated as this type
operations. However, strictly speaking, it is not possible for a replica
to verify deterministically another replica's proposal for the current
clock value without imposing stronger restriction on the synchrony
of the distributed system (e.g., bounds on message propagation and
request execution). 

\item {\em Non-verifiable pre-determinable nondeterminism (NPRE).} 
Online gaming applications, such as Blackjack~\cite{pokercheating}
and Texas Hold'em~\cite{ssbook}, exhibit this type of nondeterminism.
The integrity of services provided by such applications depends on the use
of good secure random number generators. For best security, it is essential
to make one's choice of a random number unpredictable, let alone verifiable
by other replicas. 

\item {\em Verifiable post-determinable nondeterminism (VPOST).} We have yet
to identify a commonly used application that exhibits this type of 
nondeterminism. We include this type for completeness.

\item {\em Non-verifiable post-determinable nondeterminism (NPOST).} 
In general, all multithreaded applications exhibit this type of nondeterminism.
For such applications, it is virtually impossible to determine which
thread ordering should be used prior to the execution of a request
without losing concurrency.
\end{itemize}

\Section{Controlling Replica Nondeterminism}
\label{ufsec}
\vspace{-0.1in}

In this section, we present core mechanisms for controlling replica
nondeterminism for Byzantine fault tolerance, and provide a brief informal
proof of the correctness of our mechanisms. Our mechanisms rely
on the same set of APIs as those in the original BFT library to
retrieve from, upload to, and verify by applications of the 
nondeterministic values, albeit with some modifications to the parameter list.
The most relevant APIs have been shown in Figure 1. Due to space limitation, 
we omit the detailed explanation of these APIs.

Our mechanisms work in the following ways. When the primary receives
a client's request, if it is ready to order the message, it invokes the 
\verb!propose_value()! callback function registered by the application layer. 
The application supplies the type of nondeterminism that would be involved 
in the execution of the request, and if applicable, the nondeterministic 
values. Depending on the type of nondeterminism returned by the application, 
the modified BFT algorithm operates differently according to the mechanisms 
described from Section~\ref{vpresec} through Section~\ref{nvpostsec}.

In practical applications, the execution of a request often involves
with more than one type of nondeterminism, for example, both
time-related nondeterminism (which is of verifiable pre-determinable
type) and multithreading-related nondeterminism (which is of
non-verifiable post-determinable type). To accommodate this
complexity, a bitmask should be used instead of an integer value to 
capture the nondeterminism type information in the 
\verb!propose_value()! and \verb!check_value()! upcalls.
However, the data structure used to store the nondeterministic values
does not need to be made more sophisticated because it is the
application's duty to generate and interpret them. Our algorithm can 
readily cope with this complexity. Using the same example, the time-related 
nondeterministic values can be determined during the pre-prepare-update phase.
The multithreading-related nondeterminism can be resolved in the post-commit 
phase.

\begin{figure*}[t]
   \includegraphics[width=6.7in]{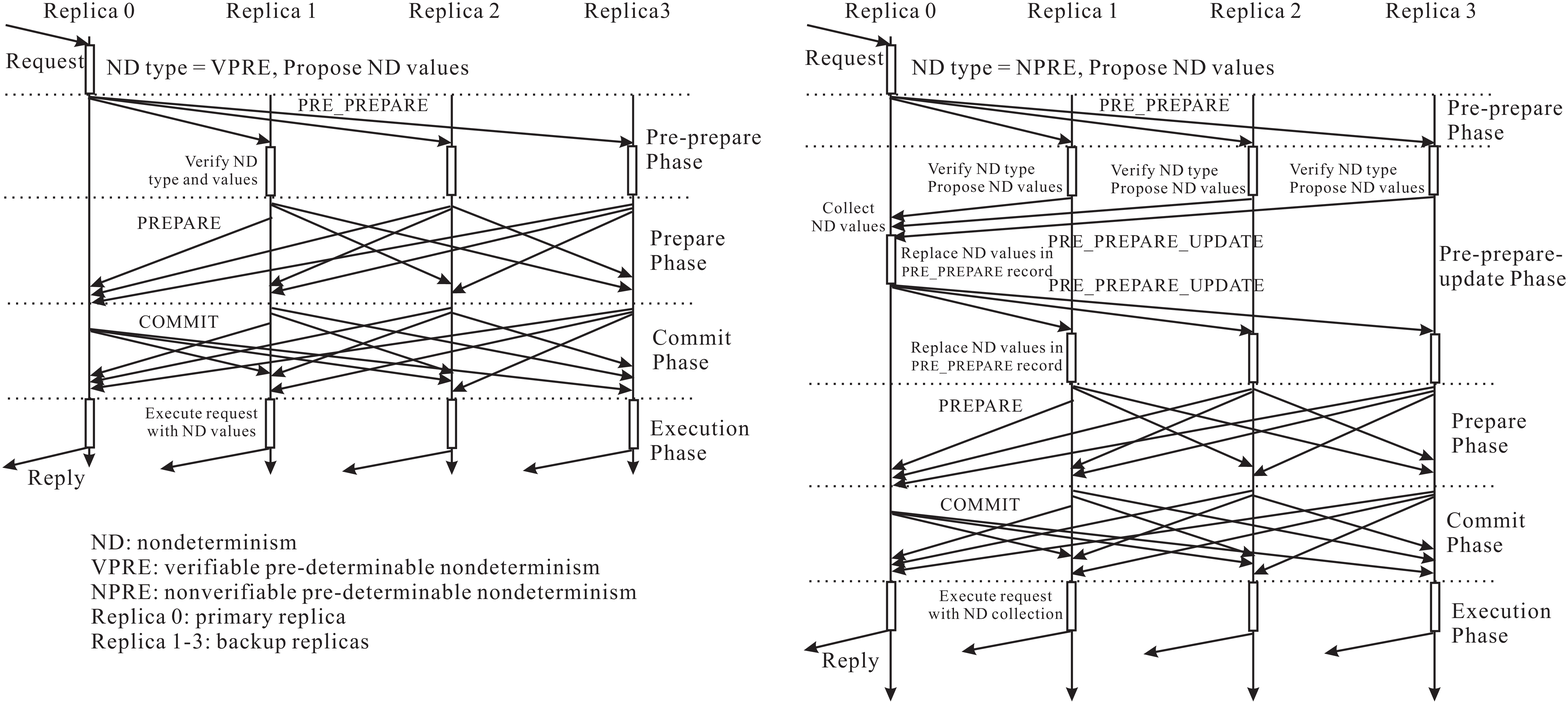}
   \caption{Normal operation of the modified BFT algorithm in handling 
verifiable (left) and nonverifiable (right) pre-determinable nondeterminism.}
   \label{prefig}
\vspace{-0.3in}
\end{figure*}

\SubSection{Controlling VPRE nondeterminism}
\label{vpresec}
\vspace{-0.1in}

If the nondeterminism for the operation at the primary is of type 
\small\verb!VPRE!\normalsize, the application
provides the nondeterministic values in the \verb!ndet! parameter. The 
obtained information is included in the 
\small\verb!PRE_PREPARE! \normalsize
message, and it is multicast to the backup replicas.

On receiving the \small\verb!PRE_PREPARE! \normalsize message, 
a backup replica invokes the \verb!check_value()! callback function. 
The replica passes the information received regarding
the nondeterminism type and data values to the
application layer so that the application can verify (1)
the type of nondeterminism for the client's request is consistent
with what is reported by the primary, and (2)
the nondeterministic values proposed by the primary is consistent
with its own values. If either check turns out to be false, 
the \verb!check_value()! call returns an error code, the backup replica 
then suspects the primary. Otherwise, the backup replica accepts 
the client's request and the
ordering information specified by the primary, logs the 
\small\verb!PRE_PREPARE! \normalsize message and multicasts a
\small\verb!PREPARE! \normalsize message to all other replicas. 
From now on, the algorithm works the same as that of the 
original BFT framework, with the exception that the 
\small\verb!PREPARE! \normalsize and \small\verb!COMMIT! \normalsize messages
also carry the digest of the nondeterministic values. The normal operations 
of the modified BFT algorithm is illustrated in Figure~\ref{prefig}.

\begin{figure*}[t]
   \includegraphics[width=6.7in]{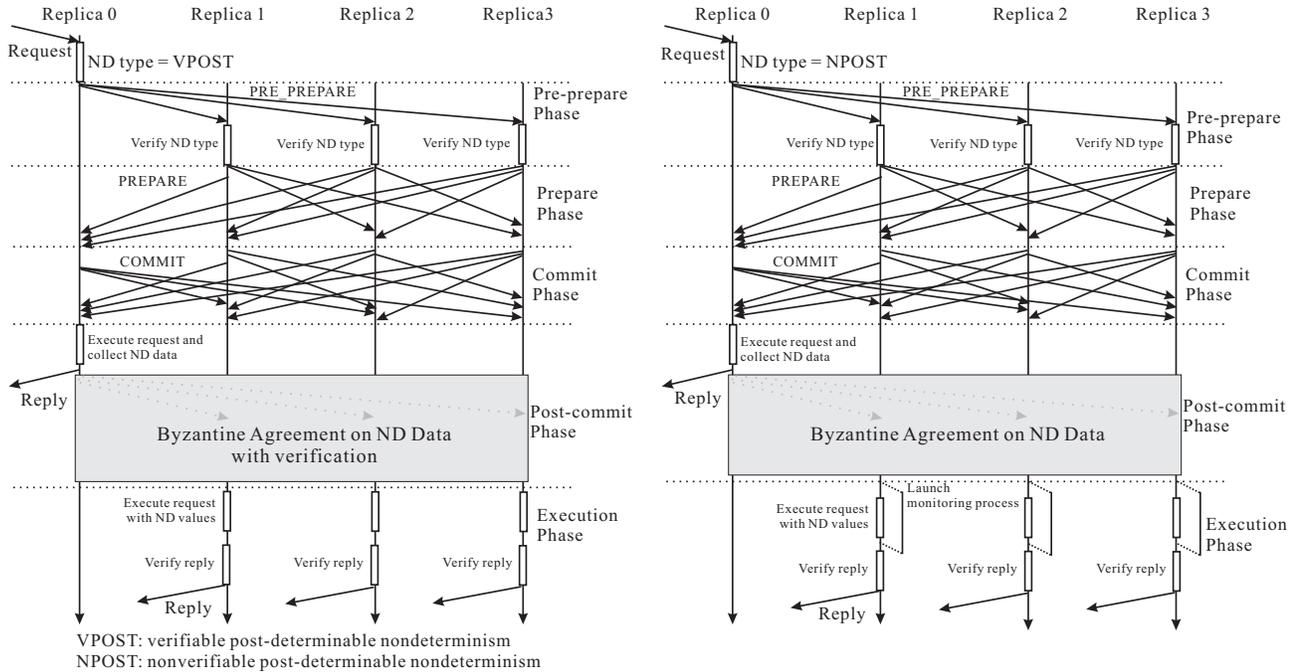}
   \caption{Normal operations of the modified BFT algorithm in handling 
verifiable (left) and nonverifiable (right) post-determinable nondeterminism.}
   \label{postfig}
\vspace{-0.3in}
\end{figure*}

\SubSection{Controlling NPRE nondeterminism}
\label{nvpresec}
\vspace{-0.1in}

If the nondeterminism for the operation at the primary is of type 
\small\verb!NPRE!\normalsize,
the application at the primary proposes its share of 
nondeterministic values. The type of
nondeterminism and the nondeterministic values are included 
in the \small\verb!PRE_PREPARE! \normalsize
message, and it is multicast to all backup replicas.

On receiving the \small\verb!PRE_PREPARE! \normalsize message, 
a backup replica invokes the \verb!check_value()! callback function 
to verify the nondeterminism type information supplied by the primary 
replica (after it has verified the client's request and the ordering 
information). If the verification is successful, the backup replica invokes 
the \verb!propose_value()! function to obtain its share of nondeterministic 
values. It then builds a \small\verb!PRE_PREPARE_UPDATE! \normalsize message
including its own nondeterministic values, and sends the message
to the primary. 

When the primary receives 2f 
\small\verb!PRE_PREPARE_UPDATE! \normalsize messages from different
backup replicas (for the same client's request), 
it builds a \small\verb!PRE_PREPARE_UPDATE! \normalsize message, including
the 2f+1 sets of nondeterministic values, each protected by the proposer's 
digital signature or authenticator. The 
\small\verb!PRE_PREPARE_UPDATE! \normalsize message itself
is further protected by the primary's signature or authenticator.
The primary then multicasts the message to all
backup replicas. From now on, the BFT algorithm operates according
to the original algorithm, except that the 
\small\verb!PREPARE! \normalsize and \small\verb!COMMIT! \normalsize messages
also carry the digest of the nondeterministic values, and the 2f+1 sets of 
nondeterministic values are delivered to the application layer as part of the 
\verb!execute()! upcall. The normal operations of the modified BFT algorithm 
for this type of nondeterminism is illustrated in Figure~\ref{prefig}.

\SubSection{Controlling VPOST nondeterminism}
\label{vpostsec}
\vspace{-0.1in}

The normal operations of the modified BFT algorithm in handling this
type of replica nondeterminism is shown in Figure~\ref{postfig}.
The primary includes the nondeterminism 
type (\ie \small\verb!VPOST!\normalsize) information in 
the \small\verb!PRE_PREPARE! \normalsize message without any 
nondeterministic values and multicasts the message to the backup replicas.

On receiving the \small\verb!PRE_PREPARE! \normalsize message, 
a backup replica performs the \verb!check_value()! upcall if it 
has verified the client's request and the ordering information. 
If the backup replica confirms the type of
nondeterminism, the BFT algorithm proceeds to the commit phase as usual.
Otherwise, the backup replica suspects the primary.

When the primary is ready to deliver the request message,
it proceeds to performing the \verb!execute()! upcall and expects 
to receive both the reply message and the recorded nondeterministic values.
Once the upcall returns, the primary stores the retrieved 
post-determined nondeterministic values, together with the digest of the 
reply, into a \verb!postnd! log, and sends the reply message 
to the client. The digest of the reply is included so that a backup replica
can verify if the primary has actually used the nondeterministic values
to generate the reply. 

A post-commit phase is needed for the primary to disseminate 
the data in the \verb!postnd! log to backup replicas and for all correct 
replicas to be sure they have received the same set of values for the 
corresponding request. Unlike 
the pre-prepare-update phase for controlling \small\verb!NPRE!\normalsize,
the post-commit phase involves with all the steps needed for correct
replicas to reach an agreement on the nondeterministic values, which requires 
three rounds of message exchanges similar to those used to determine
the ordering of the requests under normal operations. 
For \small\verb!NPRE!\normalsize, the prepare and commit phases needed for 
the correct replicas to reach a Byzantine agreement on the nondeterministic 
values are integrated with those for the corresponding request message. 
We could not do so for post-determinable nondeterminism types because the 
ordering for the corresponding request has already been decided.

A backup replica does not deliver a request message until a Byzantine 
agreement has been reached on the nondeterministic values for the request.
If the Byzantine agreement could not be reached, or the verification of 
the nondeterministic values fails, a replica suspects the primary.
Furthermore, when the replica produces a reply for the request, the digest 
of the reply is compared with that supplied by the primary. If the two do not 
match, the backup replica suspects the primary. Regardless of the comparison 
result, the backup replica sends the reply message to the client. It is safe 
to do so because if all correct backup replicas produce the same reply
using the same set of nondeterministic values (even if they
might be different from the set actually used by the primary,
which implies that the primary is lying and will be suspected),
the result is valid. 

\SubSection{Controlling NPOST nondeterminism}
\label{nvpostsec}
\vspace{-0.1in}

The handling of non-verifiable post-determinable nondeterminism 
involves with the same steps as those described in the previous subsection
until a backup replica is ready to deliver the request with the 
post-determined nondeterministic values, as shown in Figure~\ref{postfig}.

The concern here is that a faulty primary could disseminate a wrong 
set of nondeterministic values hoping to either confuse the backup replicas,
or to block them from providing useful services to their clients.
For example, if the nondeterministic values contain thread ordering 
information, a faulty primary can arrange the ordering in such a way
that it leads to the crash of the backup replicas (e.g., if the 
primary knows the existence of a software bug that leads to
a segmentation fault), or it may cause a deadlock at the backup 
replicas (it is possible for a replica to perform a deadlock analysis 
before it follows the primary's ordering to prevent this from happening).

Because in general the replica cannot completely verify the correctness
of the nondeterministic values until it actually executes the request, 
it is important for a backup replica to launch a separate
monitoring process prior to invoking the \verb!execute()! call. 
Should the replica run into a deadlock or a crash failure, 
the monitoring process can restart the replica and suspect the primary.

If it can successfully complete the \verb!execute()! upcall, the backup
replica performs the same reply verification procedure as that described in
the previous subsection, and sends the reply to the client.

\SubSection{Proof of Correctness}
\vspace{-0.1in}

We now provide an informal proof of correctness of our mechanisms.
Due to space limitation, we only argue for the correctness of the
safety property of our mechanisms and omit the proof for liveness. 
Since we do not have space to elaborate the view change mechanisms, the
proof is further limited to the safety property within a single view.

{\em Theorem 1. If a correct replica delivers a request $m$ with 
a set of nondeterministic data in view $v$, then no other 
correct replica delivers $m$ with a different set of nondeterministic data,
and all such correct replicas use, or record (at the primary), the same set of 
nondeterministic data during its execution for $m$.}

For \small\verb!VPRE! \normalsize type, 
the nondeterministic data is proposed by the primary
and the agreement on the data is carried out together with the request
message itself. At the end of the three-phase BFT algorithm, if some correct 
replicas agree on the ordering of the request message, they reach an agreement
on the nondeterministic data as well. 
For \small\verb!NPRE! \normalsize type, the nondeterministic
data is collectively determined by the pre-prepare-update phase, and it is
followed by the three-phase BFT agreement. Again, if some correct replicas 
commit the request $m$, they also agree on the associated nondeterministic 
data. For both \small\verb!VPRE! \normalsize and \small\verb!NPRE! 
\normalsize types, when the request $m$ is delivered at a 
correct replica, the nondeterministic data that have been agreed-upon are 
also delivered and used for execution.

For \small\verb!VPOST! \normalsize and \small\verb!NPOST! \normalsize types, 
the agreement on the nondeterministic data among 
correct replicas are guaranteed by the three-phase BFT algorithm executed 
during the post-commit phase. When the request $m$ is delivered at a correct
backup, the nondeterministic data associated with $m$ is also delivered. The
primary, if it is correct, must have recorded the nondeterministic data
during its execution of $m$, and have disseminated the data to
the backups during the post-commit phase. Therefore, the same nondeterministic 
data are used for execution at the primary (if it is correct) and other 
correct replicas.

\Section{Implementation and Performance}
\label{implsec}
\vspace{-0.1in}

We implemented the core mechanisms described in the previous section in C++
and integrated them into the BFT 
framework~\cite{bft-osdi99, bft-osdi2000, bft-acm, base}. 
The experiments described below are focused on the evaluation of 
the cost for providing Byzantine fault tolerance to nondeterministic 
applications in the BFT layer. The cost associated with recording 
nondeterministic values, verifying such values, and replaying such values in
the application layer is not studied in this work.

The development and test platform consists of 14 nodes running RedHat 8.0
Linux. Of the 14 computers, 4 of them are equipped with Pentium-4 2.8GHz 
processors and the rest have pentium-3 1GHz processors. The computers are 
connected via a 16-port Netgear 100Mbps switch. The server replicas run on 
the four Pentium-4 nodes and the clients are distributed across the rest 
of the nodes.

\begin{figure*}[t]
\leavevmode
\hbox to \textwidth{
\vbox{
\hbox to 3.4in{\epsfxsize=3.4in \epsfbox{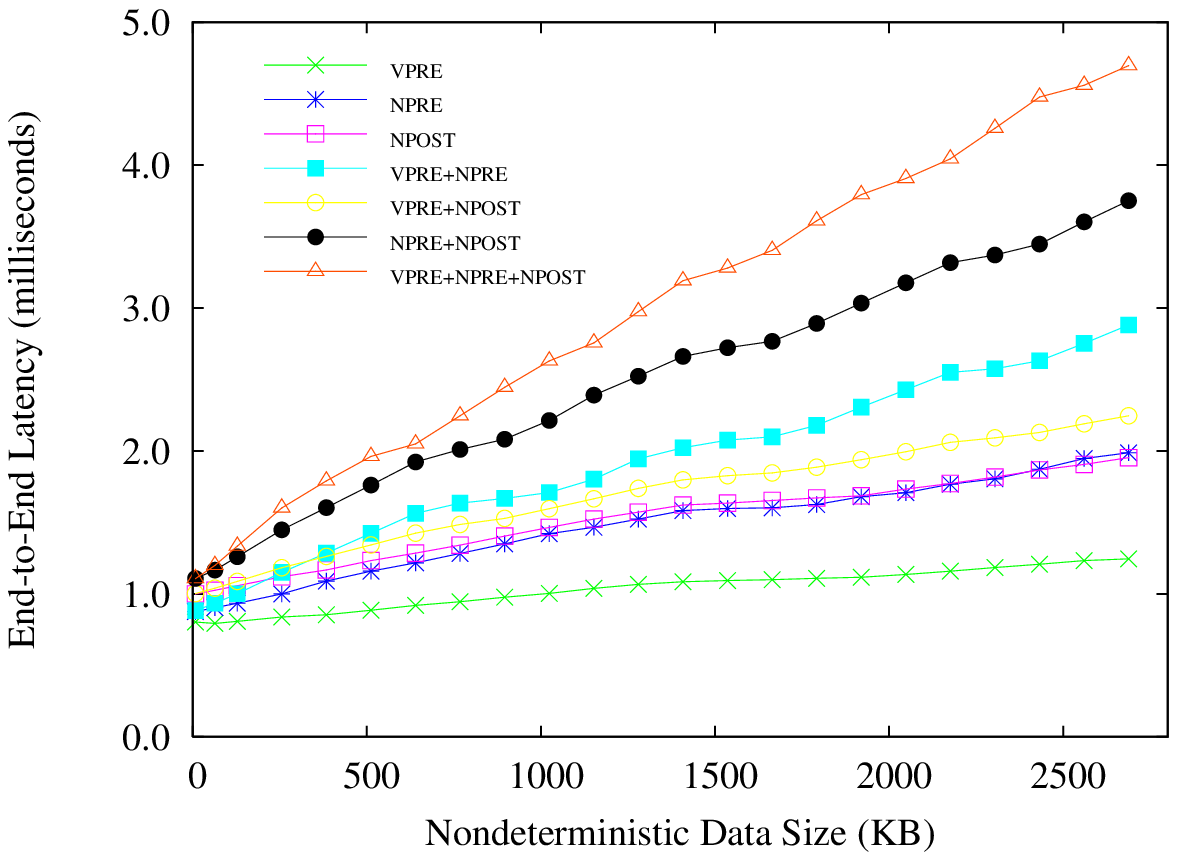}}
\hbox{\hspace{1.75in}(a)}
}
\hfil
\vbox{
\hbox to 3.4in{\epsfxsize=3.4in \epsfbox{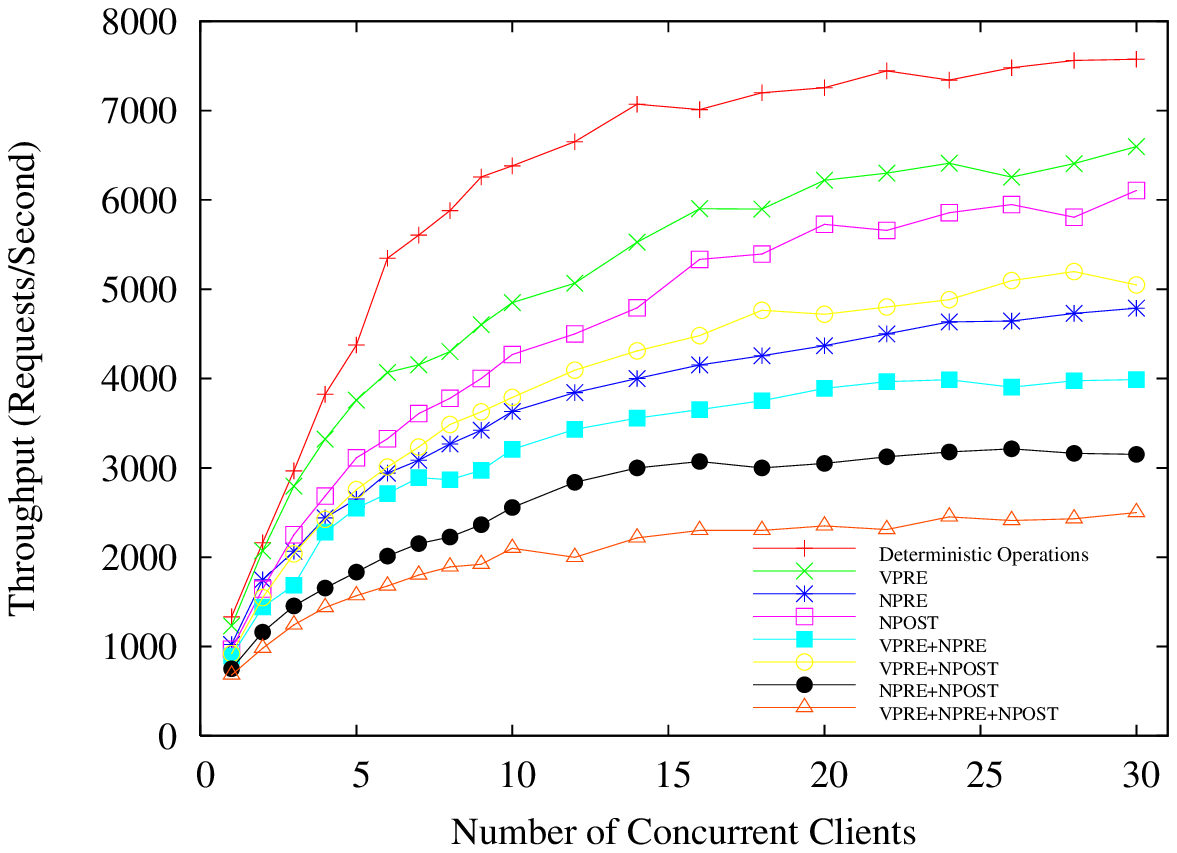}}
\hbox{\hspace{1.75in}(b)}
}}
\caption[]{End-to-end latency (a) and throughput (b) for requests with 
different types of replica nondeterminism under normal operations. In (b),
throughput for deterministic requests is included as a reference.}
\label{perffig}
\vspace{-0.25in}
\end{figure*}

Figure~\ref{perffig} shows the summary of the end-to-end latency and throughput
measurements for a client-server application under normal operations for 
different types of replica nondeterminism, including composite types. 
In each iteration, each client issues a request to the server replicas 
and waits for the corresponding reply. There is no waiting time between
consecutive iterations. The size of each request and reply is kept fixed at
1KB. In each run, we measure the total elapsed time for 10,000 consecutive 
iterations at each client. From the measured time, we derive the average 
end-to-end latency for each request-reply iteration and the system throughput.

The type of nondeterminism and the size of nondeterministic 
values vary in different experiments, except during the throughput
measurements, where the nondeterministic values are kept at 256 Bytes
for each type. Note that the sizes of nondeterministic values shown in
the horizontal axis in Figure~\ref{perffig}(a) are for {\em each} type. 
That means, for composite types, the total size of nondeterministic values 
is twice or three-times as large as those displayed.

Except for \small\verb!VPRE!\normalsize, the handling of other types of
nondeterminism involves with one or more phases of message exchanges 
for correct replicas to 
reach an agreement on the nondeterministic values. As such, as shown in 
Figure~\ref{perffig}, the end-to-end latency is noticeably larger, and the
throughput is smaller, than that of \small\verb!VPRE! \normalsize 
nondeterministic operations. The end-to-end latency difference is more 
significant as the size of nondeterministic values involved with each 
operation increases.

The results shown in Figure~\ref{perffig} are obtained after a number
of optimizations to the mechanisms described previously.
Without these optimizations, the latency is significantly larger and
the throughput is much lower, except those for \small\verb!VPRE! \normalsize
nondeterministic operations.

In the pre-prepare-update phase, which is needed to handle 
\small\verb!NPRE! \normalsize nondeterminism and other composite types
involving with \small\verb!NPRE! \normalsize nondeterminism, each backup 
replica multicasts its contribution of the nondeterministic values to all 
other replicas, and the primary decides on the collection 
(must include the contributions from 2f+1 replicas, including its 
own) to be used to calculate the final nondeterministic values. Instead of
multicasting the collection of nondeterministic values, the primary
replica disseminates the collection of the {\em digests} of the 
values proposed by each replica. This sharply reduces
the message size if the size of nondeterministic values is large.
Since each replica can log the nondeterministic values received from
other replicas, a (backup) replica can verify the digests provided
by the primary using its local copies. A backup replica might
not have received the values proposed by one or more replicas included
in the primary's message, in which case, the replica asks for retransmission
of the values.

During the post-commit phase, which is needed to handle 
\small\verb!NPOST! \normalsize nondeterminism, 
The data in the \verb!postn! log is piggybacked with the 
\small\verb!PRE_PREPARE! \normalsize message for the next request. This way,
the Byzantine agreement for the nondeterministic values is reached
together with that for the ordering of the that request, which reduces
the number of messages needed to handle this type of nondeterminism.
Even though the end-to-end latency for a request increases
slightly as a result, the system throughput is significantly improved.
To avoid waiting indefinitely for the next request, the primary
sets a timer. When the timer expires, the primary initiates the
Byzantine agreement phases for the nondeterministic values in conjunction 
with a null request so that the existing mechanisms can be reused.

It may be surprising to see that the end-to-end latency for a request
with \small\verb!NPRE! \normalsize nondeterminism is similar to, or
slightly larger than, that for a request with 
\small\verb!NPOST! \normalsize nondeterminism when there are large quantity 
of nondeterministic values. With the above optimization, the 
pre-prepare-update phase (needed to handle \small\verb!NPRE!\normalsize)
involves with at least two large messages (one message per backup replica
on its proposed nondeterministic values) while the post-commit phase 
(needed to handle \small\verb!NPOST!\normalsize) involves with only one 
large message (sent by the primary). 
Due to the same reason, the throughput for 
requests with \small\verb!NPOST! \normalsize nondeterminism is higher 
for those with \small\verb!NPRE! \normalsize nondeterminism when
sufficient number of concurrent clients are present (so that virtually
all post-determinable nondeterministic values are piggybacked with
the \small\verb!PRE_PREPARE! \normalsize messages for other requests,
rather than being sent as separate messages).

\Section{Related Work}
\label{rwsec}
\vspace{-0.1in}

Replica nondeterminism has been studied extensively under the
benign fault model \cite{Iyer:thrsched, lsa, tft, hypervisor, 
Peris:srds, pmms-patent, pmms-patent2, NMM:SRDS99, Powell:Delta4, priya06, zhao:words05}. 
However, there is a lack of systematic
classification of the common types of replica nondeterminism, and
even less so on the unified handling of such nondeterminism.
\cite{tft, hypervisor, Powell:Delta4} did provide a 
classification of some types of replica nondeterminism. However, they 
largely fall within the types of wrappable nondeterminism and verifiable 
pre-determinable nondeterminism, with the exception of nondeterminism
caused by asynchronous interrupts, which we do not address in this work.

The replica nondeterminism caused by multithreading has been 
studied separately from other types of nondeterminism, again, 
under the benign fault mode only, in \cite{Iyer:thrsched, lsa, 
Peris:srds, pmms-patent, pmms-patent2, NMM:SRDS99}. However, these
studies provided valuable insight on how to approach the problem
of ensuring consistent replication of multithreaded applications.
It is realized that what matters in achieving replica consistency
is to control the ordering of different threads on access of shared
data. The mechanisms to record and to replay such ordering have been
developed. So do those for checkpointing and restoring the state
of multithreaded applications (for example,~\cite{dieter}). 
Even though these mechanisms alone
are not sufficient to achieve Byzantine fault tolerance for multithreaded
applications, they can be adapted and used towards this goal. In this paper,
we have shown when to record and (partially) verify the ordering, 
how to propagate the ordering, and how to provision for problems encountered
when replaying the ordering, all under the Byzantine fault model.

Under the Byzantine fault model, the main effort on the subject of
replica nondeterminism control so far is to cope with
wrappable and verifiable pre-determinable replica 
nondeterminism~\cite{bft-osdi99, bft-osdi2000, bft-acm, base}. 
In \cite{bft-osdi99}, Castro and Liskov provided a brief guideline
on how to deal with the type of nondeterminism that requires collective
determination of the nondeterministic values. The guideline is
very important and useful, as we have followed in this work. However,
the guideline is applicable to only a subset of the problems we have addressed.

\vspace{-0.1in}
\Section{Conclusion and Future Work}
\label{concsec}
\vspace{-0.1in}

In this paper, we presented a classification of common types of replica
nondeterminism, and the mechanisms necessary to handle them in the
context of Byzantine fault tolerance. We also described how to integrate
our mechanisms into a well-known BFT 
framework~\cite{bft-osdi99, bft-osdi2000, bft-acm, base}.
Furthermore, we conducted extensive experiments to evaluate the performance
of the BFT framework extended with our mechanisms. We show that our
mechanisms only incur moderate runtime overhead. 

Future work will focus on the development of modules and tools that
help applications record, verify (if applicable), and replay 
nondeterministic values.

\end{document}